\newif\ifssc
 \sscfalse 

\ifssc
 \documentstyle[epsf]{elsart} 
\else
 \documentstyle[twoside,aps,prl,epsf,floats]{revtex} 
\fi

\begin{document}

\ifssc
 \begin{frontmatter} 
\else
 \twocolumn[\hsize\textwidth\columnwidth\hsize\csname 
 @twocolumnfalse\endcsname 
 \draft 
\fi

\title{Weak Localization and Negative Magnetoresistance in
Wurtzite-type Crystals}

\author{F. G. Pikus}

\address{ University of California, Santa Barbara CA 93106 USA}

\author{G. E. Pikus}

\address{A.F. Ioffe Physicotechnical Institute 194021 St  Petersburg,
RUSSIA}

\ifssc
 \begin{keyword} 
\else
 \date{\today} 
 \maketitle 
 \begin{center}{\bf Keywords:} 
\fi
A. semiconductors, D. quantum localization, electronic transport.
\ifssc
 \end{keyword} 
\else
 \end{center} 
\fi

\begin{abstract}
We have developed  a theory of  the negative magnetoresistance  due to
the weak localization in uniaxial wurtzite-type crystals, in which the
spin splitting of  the conduction band  is linear in  the wave vector,
unlike the cubic ${\rm A_3 B_5}$ crystals. Unlike earlier theories, we
take into account the correlation between the electron motion in  spin
and  co-ordinate  spaces.  It  is  shown  that  as  a  result  of this
correlation the  magnetoresistance depends  on the  orientation of the
magnetic field with respect  to the main axis  of the crystal even  if
the effective mass is isotropic.  The new theory allows to  accurately
determine  the  value  of  the  spin  splitting  constant  in uniaxial
crystals.
\end{abstract}

\ifssc
 \end{frontmatter} 
\else
 \vskip 2pc ] 
 \narrowtext 
\fi

The effect of the negative magnetoresistance observed in highly  doped
semiconductors  and  metals  in  weak  magnetic  fields is known to be
caused by the weak localization.  The weak localization, which is  the
result  of  the  constructive  interference  of  two  electron   waves
traveling along a closed path in opposite directions and scattering by
the  same  impurities,  leads  to  an  additional  contribution to the
resistance.  In  a  magnetic  field  the  two  waves  acquire  a phase
difference $\Phi/\Phi_0$,  where $\Phi$  is the  magnetic flux through
the  electron  trajectory  and  $\Phi_0  =  \hbar  c/2  e$ is the flux
quantum. As a result, the interference conditions are violated and the
additional  resistance  decreases,  which  manifests  itself as a {\em
negative  magnetoresistance},  i.e.   increase  of  the   conductivity
\cite{Altshuler80}.

If one takes into account the electron spin, the singlet state of  the
two waves with the total momentum $J=0$ gives a negative  contribution
into the resistance (antilocalization), while the triplet states  with
$J=1$  give  a  positive  contribution  \cite{Hikami80}.  In   certain
situations  the  triplet  contribution  is  suppressed  by  the   spin
relaxation,  while  the  singlet  contribution  is  not.  Then  a weak
magnetic field first reduces the negative contribution of the  singlet
state, which  results in  a positive  magnetoresistance. The  negative
magnetoresistance  takes  over  in  larger  fields, which suppress the
triplet states stronger  than the spin  relaxation does. According  to
\cite{Hikami80,Altshuler81}    this    change    of    the   sign   of
magnetoresistance in bulk crystals can be observed for  Elliott--Yafet
and Dyakonov--Perel mechanisms  of the spin  relaxation. On the  other
hand, the spin relaxation dominated by the scattering on  paramagnetic
impurities   suppresses    the   singlet    contribution,   and    the
magnetoresistance is negative  in the whole  range of magnetic  fields
where the weak localization is  important. Therefore, the type of  the
magnetoresistance  is  determined  by  the  dominant  spin  relaxation
mechanism.  In  the   non-centrosymmetric  semiconductors  (with   the
exception  of  narrow-gap  and  gapless  semiconductors)  this  is the
Dyakonov--Perel mechanism, which  is caused by  the spin splitting  of
the conduction band \cite{Dyakonov71,Pikus84}.

In  the  uniaxial  wurtzite-type  crystals,  unlike the cubic crystals
${\rm A}_3 {\rm  B}_5$, the spin  splitting of the  conduction band is
linear in  the wave  vector $\vec{k}$,  as described  by the following
Hamiltonian \cite{Rashba61}:

\begin{equation}
{\cal H} = \alpha [\vec{\sigma} \vec{k}]_z.
\label{H}
\end{equation}

\noindent where  $z$ is  the main  axis $C_6$  of the  crystal. It was
shown \cite{Iordanskii94,Pikus95,Knap96} that in this case the  theory
of  the  weak  localization  must  take  into  account the correlation
between the electron motion in co-ordinate and spin spaces,  similarly
to the $2D$ structures.

The  weak  localization  contribution  to  the  conductivity  can   be
expressed through  the Cooperon  $ {{\rm  \kern.24em \vrule width.05em
height1.4ex    depth-.05ex    \kern-.26em    C}}    $    as    follows
\cite{Altshuler80,Hikami80,Altshuler81}:

\begin{equation}
\Delta\sigma = - 2e^2 D \nu_0 \tau_0^2 \sum_{\alpha, \beta} \int_{q <
q_{\rm max}}
{{\rm  \kern.24em \vrule width.05em
height1.4ex depth-.05ex  \kern-.26em C}}
_{\alpha \beta \beta \alpha}(\vec{q}) \frac{d^3 q}{(2 \pi)^3}.
\label{sigma}
\end{equation}

\noindent  Here   $D  =   \frac{1}{3}v^2  \tau_1$   is  the  diffusion
coefficient, $\tau_1$ is the transport time, $\tau_0$ is the  momentum
lifetime, $v$ is the  electron velocity, $q_{\rm max}^2  = 1/D\tau_1$,
\\ $\nu_0 = (2 m)^{3/2} \sqrt{\epsilon_F}/(2 \pi)^2$ is the density of
states at the Fermi level $\epsilon_F$, and $m$ is the effective mass.
The  summation  in  Eq.~(\ref{sigma})  is  carried  over  spin indices
$\alpha{\rm,} \beta = \pm 1/2$.\footnote{Throughout this paper we take
$\hbar = 1$ except in the final expressions}. For simplicity we assume
an  isotropic  electron  spectrum.  Then  the  matrix equation for the
Cooperon     can      be     written      similarly     to      Refs.~
\cite{Iordanskii94,Pikus95,Knap96}:

\begin{eqnarray}
&&{{\rm  \kern.24em \vrule width.05em
height1.4ex depth-.05ex  \kern-.26em C}}_
{\vec{k}, \vec{k}^\prime}(\vec{q})
= \left|V_{\vec{k},\vec{k}^\prime}\right|^2
+ 2 \pi \nu_0 \tau_0 \int \frac{d O_g}{4 \pi}
\left|V_{\vec{k},\vec{g}}\right|^2
\nonumber \\
& & \times \Bigg\{1 - i (\vec{v}_g \vec{q}) \tau_0
- i(\vec{\sigma} + \vec{\rho})\vec{\Omega}\tau_0
- (\vec{v}_g \vec{q})^2 \tau_0^2
\label{Cooperon} \\
& & -[(\vec{\sigma}+
  \vec{\rho})\vec{\Omega}) \tau_0]^2
- 2 (\vec{v}_g \vec{q})
    (\vec{\sigma} + \vec{\rho}) \vec{\Omega}
\tau_0^2
- \frac{\tau_0}{\tau_\varphi} \Bigg\}
{{\rm  \kern.24em \vrule width.05em
height1.4ex depth-.05ex  \kern-.26em C}}_
{\vec{g}, \vec{k}^\prime}(\vec{q}).
\nonumber
\end{eqnarray}

\noindent Here, according to Eq.~(\ref{H}),

\ifssc
$$
\Omega_x(\vec{g}) = \Omega \sin\vartheta \sin\varphi, \
\Omega_y(\vec{g}) = -\Omega \sin\vartheta \cos\varphi, \
\Omega = \alpha k_F, \ \vec{v}_g = \frac{\vec{g}}{m},
$$
\else
\begin{eqnarray*}
\Omega_x(\vec{g}) &=& \Omega \sin\vartheta \sin\varphi, \\
\Omega_y(\vec{g}) &=& -\Omega \sin\vartheta \cos\varphi, \
\Omega = \alpha k_F, \ \vec{v}_g = \frac{\vec{g}}{m},
\end{eqnarray*}
\fi

\noindent   and   $\tau_\varphi$   is   the  phase-breaking  time.  In
Eq.~(\ref{Cooperon})  the  Pauli  matrices  $\vec{\sigma}$  act on the
first pair of spin indices, while the matrices $\vec{\rho}$ -- on  the
second pair.  The angles  $\vartheta$ and  $\varphi$ are  that of  the
vector $\vec{v}$ in the coordinate system with $x \parallel C_6$,  and
the integration is over solid angle $4 \pi$: $d O_g = \sin\vartheta \,
d\vartheta   \,   d\varphi$.   $V_{\vec{k},\vec{k}^\prime}$   is   the
scattering  matrix  element,  which   includes  the  density  of   the
scatterers. Then, for the scattering probability we have

$$
W(\vec{k},\vec{k}^\prime) = 2 \pi \nu_0
\left|V_{\vec{k},\vec{k}^\prime}\right|^2.
$$

If the  scattering probability  depends only  on the  scattering angle
$\theta$, then the equation

\begin{equation}
\lambda {{\rm  \kern.24em \vrule width.05em
height1.4ex depth-.05ex  \kern-.26em C}}_
{\vec{k}, \vec{k}^\prime} =
\tau_0 \int W(\vec{k}^\prime, \vec{g})
{{\rm  \kern.24em \vrule width.05em
height1.4ex depth-.05ex  \kern-.26em C}}_
{\vec{g}, \vec{k}^\prime}
\frac{d O_g}{4 \pi},
\label{Harmonics}
\end{equation}

\noindent has the following harmonics as its eigenfunctions:

\begin{equation}
{{\rm  \kern.24em \vrule width.05em
height1.4ex depth-.05ex  \kern-.26em C}}^{n,m}_
{\vec{k}, \vec{k}^\prime} =
{{\rm  \kern.24em \vrule width.05em
height1.4ex depth-.05ex  \kern-.26em C}}_{n,m}
{\rm Y}^n_m(\vartheta^\prime, \varphi^\prime).
\label{eigenfun}
\end{equation}

These formulas  are which  is written  in the  coordinate system  with
$z^\prime  \parallel   \vec{k}^\prime$,  and   $\vartheta^\prime$  and
$\phi^\prime$ are  the polar  and axial  angles of  $\vec{k}$ in  this
system.  The  following  property  is  well-known  for  the  spherical
harmonics:

\begin{eqnarray}
\int \frac{d O_g}{4 \pi}
W_{\vec{k}, \vec{g}} {\rm Y}^n_m({\bf O}_g)
&=& {\rm Y}^n_m({\bf O}_k) \int_0^{\pi/2}
W(\theta) \cos n\theta \sin \theta \, d \theta \nonumber \\
&=& {\rm Y}^n_m({\bf O}_k) \left(\frac{1}{\tau_0} -
\frac{1}{\tau_n}\right).
\label{spher}
\end{eqnarray}

\noindent Here

\begin{eqnarray}
\frac{1}{\tau_n} &=& \int_0^{\pi/2}
W(\theta) (1 - \cos n\theta) \sin \theta \, d \theta, \ (n \ne 0),
\nonumber \\
\frac{1}{\tau_0} &=& \int_0^{\pi/2}
W(\theta) \sin \theta \, d \theta.
\label{tau}
\end{eqnarray}

\noindent  it  follows  from  (\ref{Harmonics}-\ref{spher})  that  the
eigenfunctions  (\ref{eigenfun})  have  the  corresponding eigenvalues
$\lambda_{n,m} = 1 - \tau_0/\tau_n$, $(n \ne 0)$, and $\lambda_0 = 1$.
Therefore, the only  large component of  the Cooperon is  the scalar $
{{\rm \kern.24em \vrule width.05em height1.4ex depth-.05ex \kern-.26em
C}}_{0,0}  \equiv  {{\rm  \kern.24em  \vrule  width.05em   height1.4ex
depth-.05ex  \kern-.26em   C}}_0$.  Since   the  right-hand   side  of
Eq.~(\ref{Harmonics}) contains linear in  $\vec{g}$ terms, it is  also
necessary to retain the harmonics $ {{\rm \kern.24em \vrule width.05em
height1.4ex  depth-.05ex  \kern-.26em  C}}_{1,m}$.  These can be found
from  (\ref{Harmonics})  if  we  substitute  ${{\rm  \kern.24em \vrule
width.05em         height1.4ex         depth-.05ex         \kern-.26em
C}}_{\vec{k},\vec{k^\prime}}  =  {{\rm  \kern.24em  \vrule  width.05em
height1.4ex depth-.05ex \kern-.26em C}}_{1,m}$ in the right-hand  side
and  ${{\rm  \kern.24em  \vrule  width.05em  height1.4ex   depth-.05ex
\kern-.26em  C}}_{\vec{k},\vec{k^\prime}}  =  {{\rm  \kern.24em \vrule
width.05em   height1.4ex   depth-.05ex   \kern-.26em   C}}_0  +  {{\rm
\kern.24em  \vrule  width.05em  height1.4ex  depth-.05ex   \kern-.26em
C}}_{1,m}$  in  the  left-hand  side,  and  keep  only terms linear in
$\vec{g}$. Then,  using Eqs.~(\ref{spher},  \ref{tau}) we  can express
${{\rm   \kern.24em   \vrule   width.05em   height1.4ex    depth-.05ex
\kern-.26em  C}}_1$  through   ${{\rm  \kern.24em  \vrule   width.05em
height1.4ex depth-.05ex \kern-.26em C}}_0$:

\ifssc
\begin{equation}
\else
\begin{eqnarray}
\fi
{{\rm  \kern.24em \vrule width.05em
height1.4ex depth-.05ex  \kern-.26em C}}^1_{\vec{k},\vec{k^\prime}} =
\ifssc \else && \fi
\sum_m
{{\rm  \kern.24em \vrule width.05em
height1.4ex depth-.05ex  \kern-.26em C}}_{1,m} {\rm Y}^1_m(\vec{O}_k) =
\ifssc \else \nonumber \\ && \fi
- i \left( \frac{\tau_1}{\tau_0} - 1 \right)
\left[(\vec{v}_k \vec{q}) \tau_0 + (\vec{\sigma} +
\vec{\rho})\vec{\Omega} \tau_0 \right]
{{\rm  \kern.24em \vrule width.05em
height1.4ex depth-.05ex  \kern-.26em C}}_0.
\label{C1}
\ifssc
\end{equation}
\else
\end{eqnarray}
\fi

\noindent  We  now  substitute  ${{\rm  \kern.24em  \vrule  width.05em
height1.4ex  depth-.05ex  \kern-.26em  C}}  =  {{\rm \kern.24em \vrule
width.05em   height1.4ex   depth-.05ex   \kern-.26em   C}}_0  +  {{\rm
\kern.24em  \vrule  width.05em  height1.4ex  depth-.05ex   \kern-.26em
C}}_1$ to the right-hand side of Eq.~(\ref{Harmonics}), express ${{\rm
\kern.24em  \vrule  width.05em  height1.4ex  depth-.05ex   \kern-.26em
C}}_1$  through  ${{\rm   \kern.24em  \vrule  width.05em   height1.4ex
depth-.05ex  \kern-.26em  C}}_0$  using  Eq.~(\ref{C1}), and integrate
both  sides  over  $dO_k$  and  $dO_{k^\prime}$  to  cancel  out   all
components ${{\rm \kern.24em \vrule width.05em height1.4ex depth-.05ex
\kern-.26em  C}}_{n,m}$  except  ${{\rm  \kern.24em  \vrule width.05em
height1.4ex  depth-.05ex  \kern-.26em  C}}_0$.  We  then arrive to the
following equation  for the  matrix $S  = 2  \pi \nu_0  \tau_0^2 {{\rm
\kern.24em  \vrule  width.05em  height1.4ex  depth-.05ex   \kern-.26em
C}}_0$:

\begin{equation}
{\cal H} S = 1,
\end{equation}

\noindent where

\begin{eqnarray}
{\cal H} =&& D q^2 + (\vec{\sigma} + \vec{\rho})^2 \Omega^2 \tau_1 +
\nonumber \\
&&(2 D \tau_1)^{1/2} \Omega \left[(\sigma_x + \rho_x) q_y -
(\sigma_y + \rho_y) q_x\right] + \frac{1}{\tau_\varphi}.
\label{H4}
\end{eqnarray}

The Green's  function for  the equation  (\ref{H4}) can  be written as
\footnote{Note that  the formula  (\protect\ref{Green}) is  valid only
when  the  operator  ${\cal  H}$  is  hermitian.  It  is,   therefore,
inapplicable, for example, when  the Zeeman splitting in  the magnetic
field is taken into account.}

\begin{equation}
S^{\alpha \gamma}_{\beta \delta}(\vec{q}) =
\sum_{r = 1}^4 \frac{1}{E_r(\vec{q})} \Psi^r_{\alpha\beta}(\vec{q})
{\Psi^r_{\gamma\delta}}^*(\vec{q}),
\label{Green}
\end{equation}

\noindent where $\Psi^r_{\alpha\beta}(q)$  and $E_r(\vec{q})$ and  the
eigenfunctions and eigenvalues of the operator ${\cal H}$:

\begin{equation}
{\cal H} \Psi^r(q) = E_r \Psi^r(q).
\end{equation}

The  basis  of  eigenfunctions  $\Psi^r$  can  be  converted  by   the
appropriate choice of their linear combinations to a basis of  singlet
and triplet states. The Hamiltonian (\ref{H4}) is then split into  two
matrices, since it has no non-zero matrix element between singlet  and
triplet functions. For the singlet state with $J=0$ and  antisymmetric
wave function $\Psi^0$ the Hamiltonian is simple:

\begin{equation}
{\cal H}_0(q) = D q^2 + \frac{1}{\tau_\varphi},
\label{H0}
\end{equation}

\noindent while for the triplet state with $J = 1$ and symmetric  wave
functions $\Psi^1_m$ ($m = 0$, $\pm 1$) it is convenient to  introduce
the momentum  operator $\vec{J}  = (\vec{\sigma}  + \vec{\rho})/2$ and
write the Hamiltonian in this form:

\ifssc
\begin{equation}
\else
\begin{eqnarray}
\fi
{\cal H}_1(q) =
\ifssc \else && \fi
D q^2 + 2 \left(2 - J_z^2 \right) \Omega^2 \tau_1
\ifssc \else \nonumber \\ && \fi
+ 2 i (D \tau_1)^{1/2} \Omega \left(J_+ q_- - J_- q_+ \right) +
\frac{1}{\tau_\varphi},
\label{H1}
\ifssc
\end{equation}
\else
\end{eqnarray}
\fi

\noindent  where  $$q_\pm  =  q_x  \pm  i  q_y  {\ \rm and \ } J_\pm =
\frac{1}{\sqrt{2}}\left(J_x \pm J_y\right).$$

\noindent  Since  the  eigenfunctions  $\Psi_r$  are  orthonormal,  it
follows from Eq.~(\ref{Green}) that

\begin{equation}
S(\vec{q}) = -\frac{1}{E_0(\vec{q})} +
\sum_{m = -1}^1 \frac{1}{E^1_m(\vec{q})}.
\label{SEig}
\end{equation}

\noindent Due to  the anisotropy of  the spin splitting  (\ref{H}) the
Hamiltonian ${\cal  H}_1$ in  a magnetic  field depends  on the  field
direction.  In  a  magnetic  field  $B  \parallel  z$  we  can use the
commutation relation

$$
\frac{i}{2} \left[q_x q_y\right] =
\left[q_+ q_-\right] = \frac{\delta}{D},
$$

\noindent where

\begin{equation}
\delta = \frac{4 e B D}{\hbar c},
\label{delta}
\end{equation}

\noindent to introduce new operators

\begin{equation}
a = \left(\frac{D}{\delta}\right)^{1/2} q_+, \
a^\dagger = \left(\frac{D}{\delta}\right)^{1/2} q_-.
\label{qz}
\end{equation}

\noindent The Hamiltonian (\ref{H0}, \ref{H1}) can be then written as

\begin{eqnarray}
{\cal H}_0(q_z) &=& D q_z^2 + \delta \{a
a^\dagger\} + \frac{1}{\tau_\varphi},
\label{H0a} \\
{\cal H}_1(q_z) &=& D q_z^2 +
\delta \{a a^\dagger\} + \frac{1}{\tau_\varphi} +
2 \left(2 - J_z^2 \right) \Omega^2 \tau_1
\nonumber \\
&&+ 2 i (D \tau_1)^{1/2} \Omega \left(J_+ a^\dagger - J_- a \right),
\label{H1a}
\end{eqnarray}

\noindent In  the basis  of the  eigenfunctions of  the operator  $\{a
a^\dagger\} = {1 \over 2} (a a^\dagger + a^\dagger a)$ these operators
have the following non-zero matrix elements

\begin{eqnarray}
\left\langle n-1 \right| a \left| n \right\rangle & = &
\left\langle n \right| a^\dagger \left| n-1 \right\rangle = \sqrt{n},
\nonumber \\
\left\langle n \right| \{a a^\dagger\} \left| n \right\rangle
& = & n + {1 \over 2}.
\end{eqnarray}

According to  Eqs.~(\ref{sigma}), (\ref{SEig})  the weak  localization
correction to the conductivity can be written as

\begin{equation}
\Delta \sigma(B) = - \frac{e^2 \delta}{8 \pi^3}
\int_{-q_{\rm max}}^{-q_{\rm max}} d q_z S(q_z),
\label{dsigma}
\end{equation}

\noindent where

\begin{equation}
S(q_z) = \sum_{n=0}^{n_{\rm max}}
\left(-\frac{1}{E^0_n(q_z)} + \sum_{m=-1}^1\frac{1}{E^m_n(q_z)}
\right),
\end{equation}

\noindent $E^m_n(q_z)$ are the eigenvalues of ${\cal H}_0$ and  ${\cal
H}_1$ and $n_{\rm  max} = 1/\delta\tau_1$.  For $B \parallel  z$ these
eigenvalues  can  be  calculated  analytically,  similarly  to   Ref.~
\cite{Iordanskii94}. Since we are only interested in the change of the
conductivity with the magnetic field, we subtract $\Delta \sigma(B=0)$
from  Eq.~(\ref{dsigma})  to  obtain  the  magnetoconductivity $\delta
\sigma(B) = \Delta \sigma(B) - \Delta \sigma(0)$. In $\delta  \sigma$,
following Ref.~ \cite{Kawabata80}, we can extend both the  integration
over $q_z$ and summation over $n$ to infinity:

\begin{eqnarray}
\delta\sigma(B) = -\frac{e^2}{4 \pi^2 \hbar l}
 && \left[  \frac{2}{\pi} \int_0^\infty F(x) \, dx - 2
f_3\left(\frac{H_\varphi}{B} + \frac{H_{\rm
SO}}{B}\right) \right. \nonumber \\
&& \left. -
f_3\left(\frac{H_\varphi}{B} + 2\frac{H_{\rm
SO}}{B}\right) +
f_3\left(\frac{H_\varphi}{B}
\right) \right],
\label{magcondz}
\end{eqnarray}

\noindent where

\begin{eqnarray}
F(x) & = & \frac{1}{a_0} + \frac{2a_0 + 1 + b}{a_1 (a_0 + b) - 2 b}
\nonumber \\
& & + \sum_{n=1}^{\infty}
\Biggl(
\frac{3a_n^2 + 2 a_n b - 1 - 2(2n + 1)b}
{(a_n + b)(a_n^2 - 1) - 2 b[(2n+1)a_n - 1]}
\nonumber \\
&& - \frac{2}{a_{n-1}} - \frac{1}{a_{n-1} + b}
\Biggr),
\nonumber \\
a_n(x) & = & n + \frac{1}{2} + \frac{H_\varphi}{B} + \frac{H_{\rm
SO}}{B} + x^2, \ b = \frac{H_{\rm SO}}{B}, \ x = q_z l,
\label{Fx}\\
H_\varphi & = & \frac{c \hbar}{4 e D \tau_\varphi}, \
H_{\rm SO} = \frac{\Omega^2 \tau_1 c \hbar}{2 e D} =
\frac{3 \alpha^2 m^2 c}{2 e \hbar^3}, \nonumber \\
f_3(z) & = & \sum_{n=0}^\infty \left\{
\frac{2}{\sqrt{n + 1 + z} + \sqrt{n + z}} -
\frac{1}{\sqrt{n + \frac{1}{2} + z}}\right\}, \nonumber
\end{eqnarray}

\noindent and $l = (c \hbar/e B)^{1/2}$ is a magnetic length.

For $\vec{B} \perp z$ we can choose axis $x \parallel \vec{B}$ we  can
introduce $a$ and $a^\dagger$ similarly to Eq.~(\ref{qz}):

\begin{equation}
D (q_y^2 + q_z^2) = \delta \{a a^\dagger\}, \
2 D^{1/2} q_y = i \delta^{1/2} (a^\dagger - a).
\end{equation}

\noindent Then the Hamiltonian (\ref{H1}) becomes

\begin{eqnarray}
{\cal H}_1(q_x) = &&D q_x^2 +
\delta \{a a^\dagger\} + \frac{1}{\tau_\varphi} +
i (2 \delta \tau_1)^{1/2} \Omega (a^\dagger - a) J_x -
\nonumber \\
&&2 (2 D \tau_1)^{1/2} \Omega q_x J_y.
\label{H1b}
\end{eqnarray}

\noindent  In  this  case  the  eigenvalues  of the Hamiltonian can be
computed only numerically, restricting the matrix size to some  large,
but  finite,  value.  It  is  convenient  to  express  the  sum of the
reciprocal eigenvalues $E^m_n(q_x)$ through the minors $|D_{ii}(q_x)|$
of the diagonal elements ${\cal H}_{1_{ii}}$ and the determinant $|D|$
of ${\cal H}_1$ as follows:

\ifssc
\begin{equation}
\else
\begin{eqnarray}
\fi
\sum_n \sum_{m = -1}^1 \frac{1}{E^m_n(q_x)} =
\ifssc \else && \fi
\sum_i \left(\frac{|D_{ii}(q_x)|}{|D|} - {\cal H}_{1_{ii}}(q_x)
\right) +
\ifssc \else \nonumber \\ && \fi
\sum_i {\cal H}_{1_{ii}}(q_x).
\label{magcondx}
\ifssc
\end{equation}
\else
\end{eqnarray}
\fi

\noindent The first sum converges rapidly for $i \rightarrow  \infty$.
The summation in the second sum can also be extended to infinity if we
subtract its value at $B = 0$ similarly to Eq.~(\ref{magcondz}).

If we omit the linear in $\Omega$ and $q$ (or $a$, $a^\dagger$)  terms
in   Eqs.~(\ref{H1a})   and   (\ref{H1b}),   the   first   terms    in
Eqs.~(\ref{magcondz})  and  (\ref{magcondx})  also  vanish.  In   this
approximation the magnetoconductivity $\delta \sigma$ does not  depend
on the orientation of the magnetic field and is given by this formula:

\begin{eqnarray}
\delta\sigma(B) = &&\frac{e^2}{4 \pi^2 \hbar l}
\left[
2 f_3\left(\frac{H_\varphi}{B} + \frac{H_{\rm
SO}}{B}\right) \right.
\nonumber \\
&& \left.  + f_3\left(\frac{H_\varphi}{B} + 2\frac{H_{\rm
SO}}{B}\right) -
f_3\left(\frac{H_\varphi}{B}
\right)\right].
\label{magcond0}
\end{eqnarray}

\noindent  This  formula  differs  from  the  one  derived  in   Ref.~
\cite{Altshuler81} and used in, for example, Ref.~ \cite{Sawicki86} in
the  value  of  $H_{\rm  SO}$:  instead  of  $H_{\rm  SO}  =  1/\delta
{\tau_s}_{xx}$    with    $1/{\tau_s}_{xx}    =    1/{\tau_s}_{yy}   =
1/2{\tau_s}_{zz} = 2  \Omega^2 \tau_1$ \cite{Dyakonov71,Pikus84},  the
Refs.~ \cite{Altshuler81,Sawicki86}  used $1/{\tau_s}_{xx}  = \Omega^2
\tau_1$.

\begin{figure}[t]
\epsfxsize=3 in
\epsffile{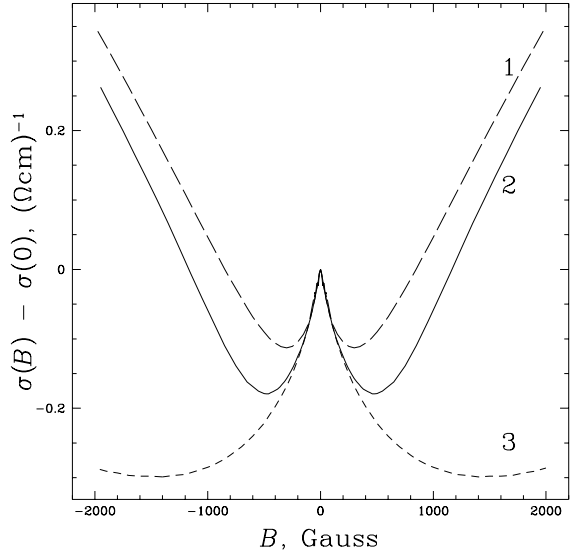}
\caption{Theoretical  magnetoconductivity  $\delta  \sigma(B)$  for $B
\perp z$ (curve 1)  and $B \parallel z$  (curve 2) for $H_\varphi  = 3
{\rm Gs}$  and $H_{\rm  SO} =  445 {\rm  Gs}$. The  curve 3  shows the
magnetoconductivity    as    obtained     from    the    theory     of
Ref.~\protect\cite{Altshuler81}  Eq.~(\protect\ref{magcond0})  for the
same values of the parameters.
\label{FigTh}}
\end{figure}

In  Fig.~\ref{FigTh}  the  magnetoconductivity  $\delta  \sigma(B)$ as
obtained   using   Eqs.~(\ref{H1a}   -   \ref{Fx}),  (\ref{H1b}),  and
(\ref{magcondx}) is shown for two orientation of the magnetic field $B
\parallel z$ and  $B \perp z$  for the same  values of $H_\varphi  = 3
{\rm Gs}$  and $H_{\rm  SO} =  445 {\rm  Gs}$. The  dashed line  shows
$\delta \sigma$ from Eq.~(\ref{magcond0})  for the same values  of the
parameters. One can see that $\delta \sigma$ significantly depends  on
the  field  direction  and   differs  strongly  from  the   result  of
Eq.~(\ref{magcond0}), which does not take into account the correlation
of the electron motion in spin and co-ordinate spaces.

\begin{figure}[t]
\epsfxsize=3 in
\epsffile{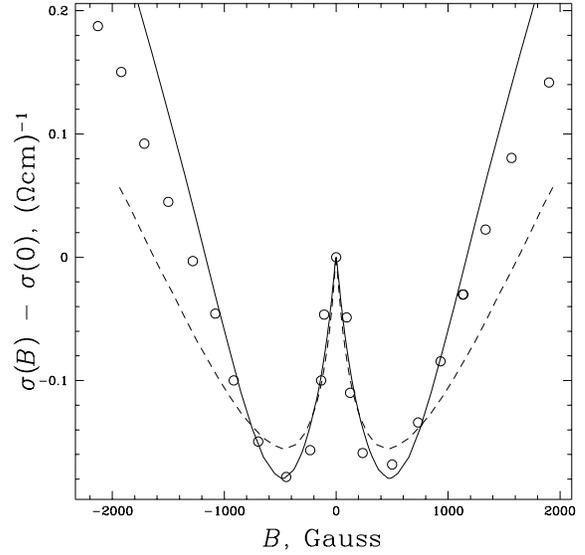}
\caption{Comparision of the measured in  Ref.~\protect\cite{Sawicki86}
(circles)  and  theoretical  magnetoconductivity.  Solid line show the
result of our theory for $H_\varphi = 3 \ {\rm Gs}$ and $H_{\rm SO}  =
445 \ {\rm  Gs}$; dashed line  is the best  fit fit obtained  from the
theory  of  Ref.~\protect\cite{Altshuler81}  ($H_\varphi  = 1.7 \ {\rm
Gs}$ and $H_{\rm SO} = 141 \ {\rm Gs}$).
\label{FigExp}}
\end{figure}

The  results  of   magnetoconductivity  measurements  for   CdSe  were
published  in  Ref.~\cite{Sawicki86}.  The  samples  used had electron
density $n = 8.6\times  10^{17} {\ \rm cm^{-3}}$  and $k_F l =  2.62$,
which corresponds to  $D = \frac{1}{3}  \frac{\hbar}{m} k_F l  = 7.84\
\frac{\rm cm^2}{\rm sec}$ for $m =  0.13 \ m_0$, $m_0$ being the  free
electron  mass.  In  Fig.~\ref{FigExp}  we  reproduce the experimental
results of Ref.~\cite{Sawicki86} for $\delta  \sigma$ at $T = 0.06  {\
\rm  K}$  and  compare  it   with  the  results  of  our   theory.  In
Ref.~\cite{Sawicki86}  the  direction  of  the  magnetic  field is not
specified; it is noted that at $T = 1.5 {\ \rm K}$ $\delta  \sigma(B)$
does  not  depend  on  the  field  orientation  (at  this  temperature
$H_\varphi \ge H_{\rm SO}$ and the dependence $\delta \sigma(B)$  does
not have a minimum at small $B$,  and the role of the linear terms  in
Eqs.~(\ref{H1a}), (\ref{H1b}) becomes  non-essential). The solid  line
in   Fig.~\ref{FigExp}    shows   the    fit   by    the   theoretical
magnetoconductivity {\bf  (}Eqs.~(\ref{dsigma} -  \ref{Fx}){\bf)}. for
$B \parallel z$. The fitting was done by weighted explicit  orthogonal
distance regression using the software package ODRPACK \cite{ODR}. The
weights were selected to increase the importance of the low-field  ($B
\le 1 kGs$) part of  the magnetoconductivity curve. The parameters  of
the best fit are $H_\varphi  = 3 \ {\rm Gs}$  and $H_{\rm SO} = 445  \
{\rm Gs}$ (these are the same values we used in Fig.~\ref{FigTh}). The
dashed line shows the best fit obtained from Eq.~(\ref{magcond0}), the
parameters of this fit  are $H_\varphi = 1.7  \ {\rm Gs}$ and  $H_{\rm
SO} = 141 \ {\rm Gs}$. One can see that the agreement with  experiment
is not as good as that of the solid curve.

The  above  value  of  $H_{\rm  SO}  =  445 \ {\rm Gs}$ corresponds to
$\alpha  =  3.9  \times  10^{-2}  \  {\rm eV \cdot \AA}$. The standard
deviation of $\alpha$ as computed by the fitting procedure is 3\%.  In
Ref.~\cite{Sawicki86}  authors  have  obtained  $\alpha  =  2.5 \times
10^{-2} \ {\rm eV \cdot \AA}$ (they have used the notation $\lambda  =
2 \alpha$ for the spin splitting coefficient). Taking into account the
error in the definition  of ${\tau_s}_{xx}$ pointed above  this result
should be  divided by  $\sqrt{2}$. In  Ref.~ \cite{Dobrowolska84}  the
constant  $\lambda$  was  measured  for  ${\rm  Cd_{1-x}Mn_xSe}$ using
electric-dipole spin  resonance and  found to  be $\alpha  = \lambda/2
\approx  (3.25  \pm  1.2)  \times  10^{-2}  \  {\rm  eV  \cdot   \AA}$
(extrapolated to $x = 0$). However, for more accurate comparison  with
our theory we must know the precise orientation of the magnetic field.
We should also  note that the  weak localization theory  includes only
the corrections to  the conductivity which  are of the  first order in
$\hbar/\tau_1 \epsilon_{\rm F} = 2/k_{\rm  F} l$. For the sample  used
in Ref.~\cite{Sawicki86} $2/k_{\rm F} l  = 0.7$, which means that  the
theory should not give very accurate results.

In conclusion, we have developed a theory of the weak localization and
the negative magnetoresistance in wurtzite-type semiconductors. It  is
demonstrated that  if the  spin splitting  of the  conduction band  is
linear in the  wave vector it  is necessary to  take into account  the
correlation  between  the  electron  motion  in  co-ordinate  and spin
spaces.  This  correlation  and  the  anisotropy of the spin splitting
result in the mangetoresistance being dependent on the magnetic  field
direction even for an isotropic electron effective mass.

\ifssc
 \begin{ack} 
\fi
F.  G.  P.  acknowledges  support  by  the NSF Grant DMR 93-08011, the
Center for Quantized Electronic Structures (QUEST) of UCSB and by  the
Quantum Institute of UCSB. G. E. P. acknowledges support by RFFI Grant
96-02-17849 and by the Volkswagen Foundation.
\ifssc
 \end{ack} 
\fi

\ifssc
\else
 \end{references} 
\fi

\end{document}